\documentclass[12pt,notitlepage,tightenlines,eqsecnum,floats,nofootinbib,amsmath,amssymb,aps,prd,superscriptaddress]{revtex4-1}

\usepackage{amsmath,amssymb,amsfonts,latexsym,cancel}

\usepackage{mathtools}
\usepackage{graphicx}
\usepackage{color}
\usepackage{amsbsy}
\usepackage{hyperref}
\usepackage{tikz}

\usepackage{seqsplit}

\usepackage{comment}

\newcommand{\be}{\begin{equation}}
\newcommand{\beq}{\begin{equation}}
\newcommand{\ee}{\end{equation}}

\def\bea {\begin{eqnarray}}
\def\eea {\end{eqnarray}}

\def\f{\frac}

\def\lp{\ell_{\rm Pl}}

\def\dd{{\rm d}}

\begin{document}

\title{Numerics of Bianchi type~II and type~IX spacetimes \\ in effective loop quantum cosmology}

\author{Timothy Blackmore} \email{timothy.blackmore@unb.ca}
\affiliation{Department of Physics, University of New Brunswick, \\
Fredericton, NB, Canada E3B 5A3}

\author{Edward Wilson-Ewing\footnote[1]{Corresponding author}} \email{edward.wilson-ewing@unb.ca}
\affiliation{Department of Mathematics and Statistics, University of New Brunswick, \\
Fredericton, NB, Canada E3B 5A3}

\begin{abstract}

We numerically determine the effective loop quantum cosmology dynamics for the vacuum Bianchi type~II and type~IX spacetimes, in particular studying how the Kasner exponents evolve across the loop quantum cosmology bounce. We find that when the spatial curvature is negligible at the bounce then the Kasner exponents transform according to the same simple equation as for a Bianchi type~I spacetime in effective loop quantum cosmology, while there are departures from this transformation rule in cases where the spatial curvature is significant during the bounce. We also use high-precision numerics to compute the evolution of a Bianchi type~IX spacetime through multiple bounces and recollapses, and find indications of chaotic behaviour. Interestingly, the numerics indicate that it is during the classical recollapse, and not the loop quantum cosmology bounce, that nearby solutions diverge most strongly.

\end{abstract}

\maketitle

\section{Introduction}

Loop quantum cosmology (LQC) is based on applying the quantization procedure of canonical loop quantum gravity in a cosmological context, with the relevant symmetries of the spacetime of interest imposed before quantization \cite{Ashtekar:2011ni}. In addition to the homogeneous and isotropic spacetimes which are of considerable observational interest (for a recent review of possible observational signatures due to LQC in the cosmic microwave background see~\cite{Agullo:2023rqq}), studying the Bianchi spacetimes---that are also homogeneous but allow anisotropies---can provide a more complete understanding of LQC effects, especially in the early universe where anisotropies may have been more important.

Further, the Bianchi spacetimes are of particular interest due to the Belinski-Khalatnikov-Lifshitz (BKL) conjecture that at generic points near a space-like singularity in general relativity, spatial derivatives become small compared to timelike derivatives and as a result neighbouring points asymptotically decouple and to a good approximation each point evolves independently following the dynamics of a Bianchi spacetime \cite{Belinski:1969, Belinski:1970ew, Uggla:2003fp, Berger:2014tev}. If the BKL conjecture is correct, then if there are some cases where the BKL approximate decoupling of neighbouring points occurs before quantum gravity effects become large, it may be sufficient to know the quantum cosmology of the Bianchi spacetimes in order to obtain a good approximation of the full quantum-corrected dynamics, point by point, in such a situation.

Three of the most interesting Bianchi spacetimes are the type~I, type~II, and type~IX models. The Bianchi~I spacetime is the simplest, without any spatial curvature, while the Bianchi~IX spacetime has the most complex dynamics that is known to be chaotic in the classical theory \cite{Chernoff:1983zz, Cornish:1996yg}; the Bianchi~II spacetime can be viewed as an intermediate case, as shall be explained in more detail in Sec.~\ref{s.rev}. Each of these cosmological spacetimes has been studied in LQC, here we will consider the LQC dynamics developed in~\cite{Ashtekar:2009vc, Ashtekar:2009um, WilsonEwing:2010rh, Martin-Benito:2011fdk, Singh:2013ava}; for earlier related work see \cite{Bojowald:2003md, Bojowald:2003xf, Chiou:2007sp, Martin-Benito:2008dfr}, and for a study of the quantum dynamics of the Bianchi~IX spacetime for a different polymer quantization based on the Misner variables see \cite{Giovannetti:2019ewe}.

In this paper, we numerically determine the LQC effective dynamics for the Bianchi spacetimes, building on previous studies \cite{Chiou:2007sp, Gupt:2012vi, Corichi:2012hy, Corichi:2015ala, McNamara:2022dmf}. In particular, we compute how the three Kasner exponents (that parametrize the rate of expansion of the three directional scale factors) evolve in the LQC effective theory for the Bianchi~II and Bianchi~IX spacetimes. We also find indications that the LQC dynamics for the Bianchi~IX spacetime may be chaotic. The LQC effective equations that we consider here can be derived from the full quantum dynamics by neglecting quantum fluctuations \cite{Ashtekar:2006wn, Taveras:2008ke} and provide an excellent approximation for the dynamics of expectation values for sharply-peaked states describing homogeneous spacetimes whose volume always remains much larger than the Planck volume \cite{Rovelli:2013zaa, Bojowald:2015fla}.

The outline of the paper is as follows. A brief review of the Bianchi spacetimes is given in Sec.~\ref{s.rev}, including a summary of their dynamics according to classical general relativity, as well as in the LQC effective theory. Then, in Sec.~\ref{s.num} we describe the numerical methods we use to fix initial conditions and calculate the dynamics. The results of the numerics are presented next, for Bianchi~II in Sec.~\ref{s.bII} and for Bianchi~IX in Sec.~\ref{s.bIX}. We end with a summary and a brief discussion in Sec.~\ref{s.disc}.

\section{Bianchi spacetimes: Review}
\label{s.rev}

We start with a brief review of the Bianchi spacetimes, considering the vacuum case, and contrast their classical dynamics with the LQC effective dynamics. While our focus is on the Bianchi type~II and type~IX spacetimes, we also describe the Bianchi type~I spacetime since it provides an excellent approximation to the Bianchi~II and Bianchi~IX spacetimes when their spatial curvature is negligible.

\subsection{General Relativity}
\label{s.rev-cl}

The metric for the Bianchi~I spacetime is
\begin{equation}
    \dd s^2 = -\dd t^2 + a_1(t)^2 \, \dd x_1^2 + a_2(t)^2 \, \dd x_2^2 + a_3(t)^2 \, \dd x_3^2,
\end{equation}
where the $a_i(t)$ are the directional scale factors.  For the vacuum spacetime, the equations of motion of general relativity give
\beq
a_i(t) = t^{k_i},
\ee
where the Kasner exponents $k_i$ are constant and satisfy $\sum_i k_i = \sum_i k_i^2 = 1$. Given these two constraints, selecting one Kasner exponent fixes the other two.  A short calculation shows that this spacetime has a curvature singularity at $t=0$ that is analogous to the big-bang singularity in the FLRW spacetimes.

Defining the mean scale factor
\begin{equation}
    a= (a_1 a_2 a_3)^{1/3},
\end{equation}
a convenient way to calculate the Kasner exponents in a numerical solution is \cite{deCesare:2019suk}
\begin{equation} \label{ki}
k_i = \frac{a}{3 a_i}\bigg(\frac{\dot{a_i}}{\dot{a}}\bigg),
\end{equation} 
where the dots denote a derivative with respect to the proper time $t$. Importantly, this equation is well-defined very generally (except at $\dot a = 0$ where the expression diverges) and can be used independently of the dynamics (whether general relativity, the LQC effective theory, or some other modified gravity theory), and for any Bianchi spacetime, to define a `quasi-Kasner' exponent even in the case that $k_i$ is no longer constant.

All Bianchi spacetimes can be described in terms of three directional scale factors $a_i(t)$, the main difference between the various Bianchi models is the presence of different types of spatial curvature; for a review of all Bianchi spacetimes see, e.g., \cite{Ryan-Shepley}. The dynamics of the directional scale factors is affected by the spatial curvature in a relatively simple manner: the spatial curvature acts as a potential energy that is well approximated by a wall that the system `bounces' off of, and the scale factors evolve approximately as $a_i \approx (t-t_o)^{k_i}$ away from the potential walls.  As a result, during the time that the spatial curvature is sufficiently small the dynamics of the spacetime are to a good approximation the same as for the Bianchi~I spacetime, and the full dynamics can be approximated by a sequence of Bianchi~I solutions for the scale factors (with different Kasner exponents $k_i$ for each approximate Bianchi~I solution in the sequence) that are separated by short transitionary periods when the spatial curvature is large.  For the Bianchi~II spacetime the potential is a single wall that (in classical general relativity) the system will bounce off exactly once, while for the Bianchi~IX spacetime the potential forms a group of walls with a closed triangular shape with the system located inside the triangle---it can be verified that in the vacuum case the classical dynamics of the system will bounce off the potential walls an infinite number of times before reaching the big-bang singularity.

In more detail, the Bianchi~II metric is
\beq
\dd s^2 = -\dd t^2 + a_1(t)^2 \, \big( \dd x_1 - x_3 \dd x_2 \big)^2 + a_2(t)^2 \, \dd x_2^2 + a_3(t) \, \dd x_3^2,
\ee
and the presence of spatial curvature generates a potential $U_{\!I\!I} \propto a_1(t)^4$ that affects the dynamics of the scale factors; the potential is given explicitly in Eq.~\eqref{Bianchi_Potentials} below, and for the exact solution in general relativity for all three scale factors see~\cite{Belinski:1969}. When the scale factor $a_1(t)$ is small, then the potential is negligible and the scale factors evolve to an excellent approximation like they do in the Bianchi~I spacetime.  Since the potential grows very rapidly as the fourth power of $a_1(t)$, it can be approximated as a hard wall that the solution bounces off, after which the scale factors once again approximately evolve as in a Bianchi~I spacetime except with different Kasner exponents.  The Kasner exponents of the two Bianchi~I solutions that approximate the dynamics of the Bianchi~II spacetime on either side of this Kasner transition are related by \cite{Belinski:1969, Belinski:1970ew}
\begin{equation} \label{kasner}
       \bar{k}_1 = \frac{-k_1}{1 + 2k_1},
       \quad \bar{k}_2 = \frac{k_2 + 2k_1}{1 + 2k_1},
       \quad \bar{k}_3 = \frac{k_3 + 2k_1}{1 + 2k_1},
\end{equation}
a short calculation shows that the new Kasner exponents $\bar k_i$ after the Kasner transition also satisfy the Kasner constraints $\sum_i \bar k_i = \sum_i \bar k_i^2 = 1$.  In summary, for a vacuum Bianchi~II spacetime the scale factors initially evolve as $a_i \sim t^{k_i}$, with constant $k_i$ satisfying the Kasner constraints $\sum_i k_i = \sum_i k_i^2 = 1$, then bounce off the potential wall when $a_1(t)$ becomes sufficiently large, and finally evolve as $a_i \sim t^{\bar k_i}$ with the $\bar k_i$ determined by \eqref{kasner}.  In the following, we will call such a bounce off a potential wall a `Kasner transition'.  Since $a_1(t)$ decreases after the transition, the potential $U_{\!I\!I}$ will also decrease and always remain negligible in the future---there is exactly one Kasner transition in the entire history of a classical vacuum Bianchi~II spacetime.

The dynamics of the Bianchi~IX spacetime is significantly richer.  The spatial manifold for the Bianchi~IX spacetime is a 3-sphere, so the metric is
\beq
\dd s^2 = -\dd t^2 + \sum_{i=1}^3 a_i(t)^2 \, \sigma_i^2,
\ee
where the $\sigma_i$ are a set of 3 orthonormal forms that provide a basis for $\mathbb{S}^3$.  The spatial curvature once more affects the dynamics of the scale factors through a potential that has the form $U_{\!I\!X} \sim a_1^4 + a_2^4 + a_3^4 + a_1^2 a_2^2 + a_1^2 a_3^2 + a_2^2 a_3^2$.  This potential has the approximate shape of a pyramid with a triangular base, with the system always remaining inside the pyramid throughout all of its dynamics.  The pyramid has an upper tip located at a maximal value for $a$, and its base at $a_i=0$ corresponding to the classical singularity.

The dynamics of a vacuum Bianchi~IX spacetime have the following properties \cite{Belinski:1970ew, Misner:1969hg}. Initially, the system starts within the pyramid, far below the upper tip of the pyramid, and with the mean scale factor $a$ increasing. While the system is far from all of the potential walls, the scale factors evolve as $a_i \sim t^{k_i}$ with constant Kasner exponents; this is called a Kasner epoch.  When the solution nears a potential wall, there will be a Kasner transition as the solution bounces off the potential wall, and the Kasner exponents change.  If the Kasner transition is a bounce off one potential wall only (either the $a_1^4,~ a_2^4$ or $a_3^4$ wall), then the Kasner exponents change exactly as in \eqref{kasner}, up to cyclic permutations of the indices depending on which potential wall the solution bounced off during the Kasner transition.  If the solution is close to two walls during a Kasner transition (for example, the $a_1^4,~ a_2^4$ and $a_1^2 a_2^2$ terms in potential are all large), then numerics is needed to calculate exactly how the Kasner exponents change during the Kasner transition.  The mean scale factor will continually increase, until the solution nears the upper tip of the pyramid, at which time the spacetime undergoes a recollapse and the mean scale factor starts to decrease.  In a classical vacuum Bianchi~IX spacetime, there is exactly one recollapse; before it the mean scale factor is monotonic increasing, and after it is monotonic decreasing.  Either side of the recollapse there are infinitely many Kasner transitions as described above.  It has been shown that the evolution of the Bianchi~IX spacetime close to the singularity in general relativity is chaotic: the dynamics are strongly mixing, have non-zero topological entropy, and its basin boundaries are fractal \cite{Chernoff:1983zz, Cornish:1996yg}.

\subsection{LQC Effective Dynamics}

The quantum theory for Bianchi spacetimes in LQC has been defined for the vacuum Bianchi~I spacetime \cite{Martin-Benito:2011fdk} as well as for the Bianchi~I, II and IX models minimally coupled to a massless scalar field \cite{Ashtekar:2009vc, Ashtekar:2009um, WilsonEwing:2010rh, Singh:2013ava}; the quantum framework for vacuum Bianchi~II and IX spacetimes can be derived in a straightforward fashion by combining these results.

While some progress has been made in determining the quantum dynamics numerically for anisotropic cosmologies \cite{Pawlowski-talk}, this has proven to be challenging. (See also \cite{Martin-Benito:2009xaf, Diener:2017lde} for numerical studies of the quantum dynamics for a different loop quantization of the Bianchi~I spacetime proposed in \cite{Martin-Benito:2008dfr}; although this quantization prescription gives simpler dynamics it has some undesirable physical properties \cite{Corichi:2009pp}.) Due to the difficulty in solving the full quantum dynamics, even numerically, most studies have focused on the LQC effective dynamics which are derived by assuming quantum fluctuations are negligible.  This is a good approximation for quantum states that are sharply peaked, and whose scale factors are always much greater than the Planck length \cite{Taveras:2008ke, Rovelli:2013zaa, Bojowald:2015fla}; for these states the LQC effective dynamics provide an excellent approximation to the evolution of the expectation values of observables of interest.

The basic variables used in LQC are the densitized triad and Ashtekar-Barbero connection; for the Bianchi spacetimes the densitized triads have three non-zero components that are parametrized by $p_i$, and similarly there are three non-zero components of the extrinsic curvature $K_i$ (not to be confused with the Kasner exponent $k_i$), that appears in the Ashtekar-Barbero connection. These variables are canonically conjugate,
\beq
\{K_i, p_j\} = 8 \pi G \, \delta_{ij},
\ee
and the $p_i$ are related to the directional scale factors by the simple relations
\beq
p_1 = a_2 a_3, \quad p_2 = a_1 a_3, \quad p_3 = a_1 a_2.
\ee
In general relativity, the $K_i$ capture the time derivatives of the directional scale factors, $K_i = \dot a_i$, but this relation is a result of the classical equations of motion and it is modified in LQC.

The LQC effective dynamics are generated by the effective Hamiltonian constraint $C_H = N\mathcal{H}$, where the effective scalar constraint $\mathcal{H}$ is given by \cite{Chiou:2007sp, Ashtekar:2009vc, Ashtekar:2009um, WilsonEwing:2010rh}
\begin{align} \label{Ham_LQC}
\mathcal{H} = &
 -\frac{ \sqrt{p_1p_2p_3}}{8 \pi G \gamma^{2} \Delta} \Big( \sin\gamma\bar{\mu}_1K_1\sin\gamma\bar{\mu}_2K_2 + \sin\gamma\bar{\mu}_2K_2\sin\gamma\bar{\mu}_3K_3 + \sin\gamma\bar{\mu}_3K_3\sin\gamma\bar{\mu}_1K_1 \Big) \nonumber \\ & \quad 
+ \f{1}{\sqrt{p_1p_2p_3}} \, U_i = 0,
\end{align}
for the `K' loop quantization of the Bianchi spacetimes \cite{Singh:2013ava}. Here $\Delta \sim \lp^2$ is the LQG area gap, the smallest non-zero eigenvalue of the LQG area operator, $\gamma$ is the Barbero-Immirzi parameter, and
\begin{equation} \label{muterm}
\bar{\mu}_1 = \sqrt{\frac{p_1 \Delta}{p_2 p_3}}, \qquad
\bar{\mu}_2 = \sqrt{\frac{p_2 \Delta}{p_1 p_3}}, \qquad
\bar{\mu}_3 = \sqrt{\frac{p_3 \Delta}{p_1 p_1}}.
\end{equation}
Neglecting inverse triad corrections, the potentials $U_i$ are exactly the same in the LQC effective dynamics as they are in classical general relativity,
\begin{equation} \label{Bianchi_Potentials}
\begin{split}
    &U_{\!I} = 0, \qquad U_{\!I\!I}(p_i)= \f{1}{32 \pi G} \cdot \frac{p_2^{2} p_3^{2}}{p_1^{2}},\\ &U_{\!I\!X}(p_i)= \frac{1}{32 \pi G}\bigg(\frac{p_2^{2} p_3^{2}}{p_1^{2}} + \frac{p_1^{2} p_3^{2}}{p_2^{2}} + \frac{p_1^{2} p_2^{2}}{p_3^{2}} - 2 (p_1^2 + p_2^2 + p_3^2) \bigg),
\end{split}
\end{equation}
these potentials are related to the spatial curvature by ${}^3 \! R = - 16 \pi G \, U_i / q $, where $q = a^6 = p_1 p_2 p_3$ is the determinant of the spatial metric.  Note that the classical Hamiltonian constraint of general relativity for Bianchi spacetimes is recovered in the limit $\Delta \to 0$.

The equations of motion follow from the constraint in the usual manner (see, e.g., \cite{Ashtekar:2009vc, Singh:2013ava} for their explicit form),
\begin{equation} \label{EqnsofMotion}
\frac{\dd p_i}{\dd \tau} = - 8 \pi G ~ \frac{\delta C_H}{\delta K_i}, \qquad
\frac{\dd K_i}{\dd \tau} = \, 8 \pi G ~ \frac{\delta C_H}{\delta p_i},
\end{equation}
recalling that $C_H = N \mathcal{H}$ and
the lapse $N$ can be chosen freely, fixing the time coordinate $\tau$. These are six coupled first-order non-linear ordinary differential equations that we shall study numerically, setting $N = \sqrt{p_1 p_2 p_3}$ because this choice gives a relatively simple set of equations of motion. (Note that in Sec.~\ref{s.rev-cl} above, the different choice for the lapse of $N=1$ was used; the relation $\dd t = \sqrt{p_1 p_2 p_3} \, \dd\tau$ can be used to translate between the two different choices for the time coordinate.) These equations of motion only have singular points when $p_1=0$ for Bianchi~II spacetimes and when (at least) one of the $p_i$ vanishes for Bianchi~IX spacetimes, with such configurations each corresponding to a singular geometry; however, these points are avoided by the effective dynamics assuming the initial configuration is non-singular, as can be seen below in Secs.~\ref{s.bII} and \ref{s.bIX}. As a result, these equations of motion do not have any singular points that are reached dynamically.

When the spacetime curvature is small compared to the Planck scale, the LQC effective dynamics are essentially identical to the classical dynamics of general relativity. But in the regime where the curvature becomes sufficiently large, LQC effects become important and cause a cosmic bounce to occur, resolving the singularity that arises in the classical theory \cite{Chiou:2007sp, Ashtekar:2009vc, Ashtekar:2009um, WilsonEwing:2010rh}.  Importantly, the energy density of matter (if any matter is present) and the shear anisotropies are all bounded above by the Planck scale \cite{Corichi:2009pp, Gupt:2011jh, Singh:2011gp, Singh:2013ava, Saini:2017ipg, Saini:2017ggt}.

\begin{figure}[t]
\begin{center}
\includegraphics[width=12cm]{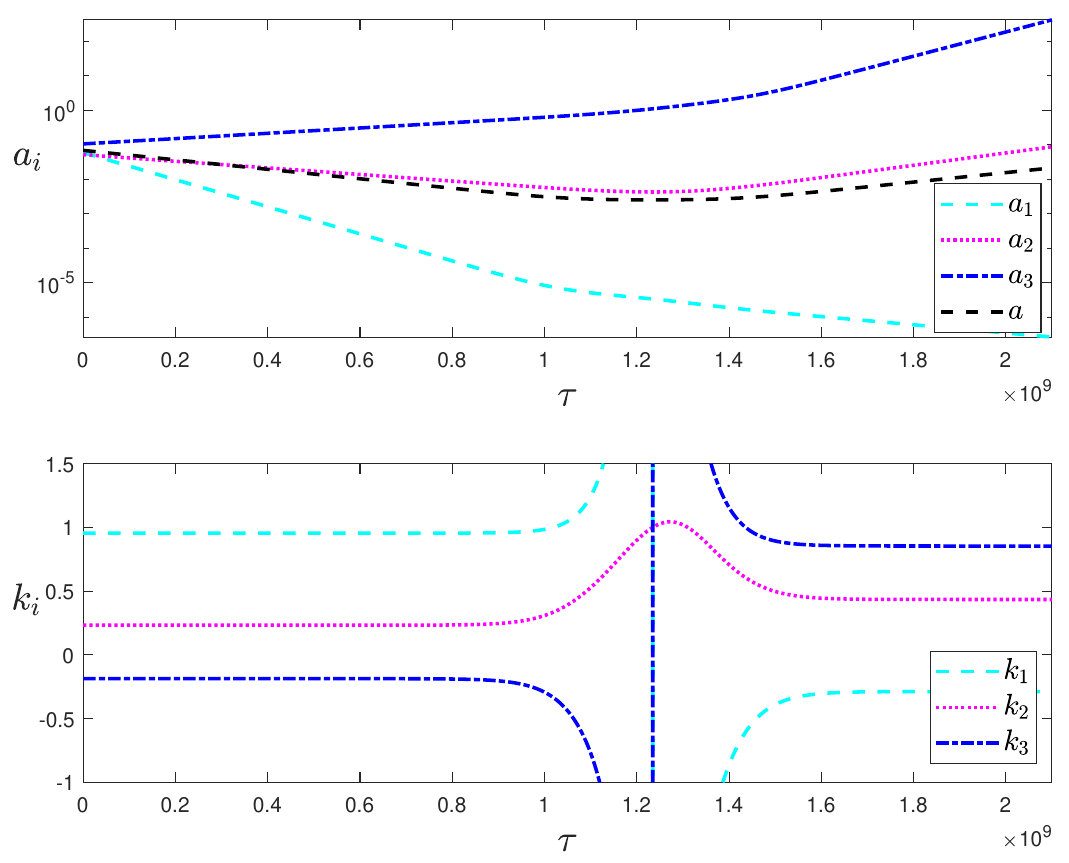}
\caption{This is a numerical solution for the LQC effective dynamics of a Bianchi~I spacetime; the top plot shows the scale factors, while the lower plot shows the Kasner exponents as calculated by Eq.~\eqref{ki}. The initial Kasner exponents $(k_1, k_2, k_3)$ are (0.954498, 0.232394, -0.186891) while the final Kasner exponents are (-0.287832, 0.434273, 0.853558). These sets of Kasner exponents before and after the LQC bounce satisfy the transformation rule \eqref{lqc-kas}.
}
\label{B1_LQC}
\end{center}
\end{figure}

Further, it has been shown that for the Bianchi~I spacetime the Kasner exponents change in a very simple fashion during the LQC bounce,
\beq \label{lqc-kas}
\bar k_i = \f{2}{3} - k_i.
\ee
This transition rule has been derived analytically \cite{Chiou:2007mg, Wilson-Ewing:2017vju} and has been verified numerically \cite{Corichi:2012hy}; a representative example is shown in Fig.~\ref{B1_LQC}.  Interestingly, in addition to LQC, this transition rule holds for a wide variety of modified gravity theories which generate a bounce that replaces the singularity of general relativity \cite{deCesare:2019suk}.

While this transition rule for the Kasner exponents at the LQC bounce is exact for the Bianchi~I spacetime, it is expected that it will also hold approximately in other Bianchi spacetimes so long as the spatial curvature is negligible during the LQC bounce.  For Bianchi~II spacetimes, there are three possibilities \cite{Wilson-Ewing:2017vju}: (i) a Kasner transition will occur before the LQC bounce but not after, (ii) a Kasner transition will occur after the LQC bounce but not before, (iii) two Kasner transitions will occur, once before the LQC bounce and one after.  As usual, the dynamics for the Bianchi~IX spacetime are richer.  The main consequence of the LQC effects is to, in effect, add a non-singular bottom to the pyramid-shaped potential $U_{\!I\!X}$ that the system bounces off \cite{Wilson-Ewing:2018lyx} (before the system would reach the classical singularity).  As a result, the LQC effective dynamics for the Bianchi~IX spacetime now consist of a sequence of bounces (due to LQC) and recollapses (due to the spatial curvature) in the mean scale factor, with each cycle interspersed with a finite number of Kasner epochs (each epoch separated by a Kasner transition).

The aim of this paper is to numerically calculate the LQC effective dynamics of the Bianchi~II and Bianchi~IX spacetimes, in particular to verify that the transition rule \eqref{lqc-kas} holds when the spatial curvature is negligible, and to determine how the Kasner exponents change during the LQC bounce in situations where the spatial curvature cannot be neglected during the LQC bounce.  We also find some indications for the presence of chaos in the Bianchi~IX spacetime, as we discuss in more detail below.

\section{Numerical Methods}
\label{s.num}

The reader who is not interested in the details of the numerics can skip this section and go directly to Secs.~\ref{s.bII} and \ref{s.bIX} without any loss of continuity. There are two main parts to the numerics: first, selecting initial conditions of interest; and second, determining the dynamics.  As is explained in more detail in the following, we use a fixed timestep sixth-order Runge-Kutta solver written in C++ for the Bianchi~II spacetime, and a high precision, variable timestep fourth-order Runge-Kutta solver written in Julia with 128-bit precision for the Bianchi~IX spacetime.  In both cases, the numerical error is estimated by the relative departure of the effective LQC Hamiltonian constraint from zero, see Eq.~\eqref{Rel_Error} below for details.  For all numerics and the remainder of this paper, we set $G = 1$,  $\gamma = 1$, $\ell_{Pl} = 1$, and $\Delta = 4 \sqrt{3} \pi$.

\subsection{Initial Conditions}

It is useful to introduce the notation
 \begin{equation}
    C_{Hi} = -\frac{p_1p_2p_3}{8 \pi G \gamma^{2} \Delta} \, \sin \gamma \bar{\mu}_jK_j \, \sin \gamma\bar{\mu}_kK_k,
\end{equation}
where $j \neq k$ are both different from $i$, so the effective LQC Bianchi~II and Bianchi~IX Hamiltonian constraints can be rewritten as
\begin{equation}
    C_H = C_{H1} + C_{H2} + C_{H3} + U_{i}.
\end{equation}

For convenience, it is simplest to choose initial conditions at an instant where both the spatial curvature (encoded in $U_{\!I\!I}$ for Bianchi~II spacetimes and encoded in $U_{\!I\!X}$ for Bianchi~IX spacetimes) as well as LQC effects are entirely negligible.  This choice entails no loss of generality since such a time always exists for Bianchi~II and IX spacetimes in the LQC effective theory (for Bianchi~II spacetimes, it is sufficient to go far in the past before any Kasner transition or LQC bounce occurs, while for the Bianchi~IX spacetime it is always possible to find a Kasner epoch between an LQC bounce and a recollapse where this is true).

To further simplify setting the initial conditions, we also assume that at the instant the initial conditions are imposed, all of the $p_i$ are approximately of the same order of magnitude, and the $K_i$ are also, among themselves, approximately of the same order of magnitude,
\begin{equation} \label{same_mag}
p_1 \sim p_2 \sim p_3 \sim p, \qquad |K_1| \sim |K_2| \sim |K_3| \sim K.    
\end{equation}
Note that we are not assuming that $p$ and $K$ are of the same order of magnitude.  This assumption implies that $C_{H1} \sim C_{H2} \sim C_{H3}$ and, for the Bianchi~IX spacetime, that all terms in $U_{\!I\!X}$ are approximately of the same order, with $U_{\!I\!X} \sim p^2$.  Note that the condition \eqref{same_mag} is only assumed to hold at $\tau = \tau_i$, the initial time when the initial conditions are set.

LQC effects are negligible when the terms in the Hamiltonian constraint are approximately equal to their classical counterparts, namely (up to a factor of $8\pi$) that $C_{Hi} \approx - p_j p_k K_j K_k$, which implies
\begin{equation} \label{part_relation}
\f{|K|}{\sqrt p} \ll 1.
\end{equation}
Next, the spatial curvature is negligible if $|C_{Hi}| \gg U_i$, or equivalently, after using \eqref{part_relation} and dropping numerical prefactors of order unity,
\begin{equation} \label{whole_relation}
K^2 \gg 1.
\end{equation}
Combining these conditions gives the hierarchy
\begin{equation} \label{hierarchy}
1 \ll K \ll p
\end{equation}
for initial conditions where both LQC effects and the spatial curvature are initially negligible in the Bianchi~II and Bianchi~IX spacetimes.

The initial conditions must satisfy the Hamiltonian constraint $C_H = 0$, so only five of the $p_i, K_i$ can be chosen freely, satisfying the hierarchy \eqref{hierarchy}.  For concreteness, we specify all $p_i$ as well as $K_1, K_2$, and then determine $K_3$ from the requirement that $C_H=0$.

In some cases, it may be necessary to choose a specific set of Kasner exponents; for example by fixing $k_1$ (due to the Kasner constraints $\sum_i k_i = \sum_i k_i^2 = 1$, once one Kasner exponent is fixed, then so are the other two).  This can be done in an iterative process, by repeatedly modifying the values of $K_1$ and $K_3$.  The idea is to first choose arbitrary $p_1, p_2, p_3, K_1, K_2$ satisfying the hierarchy \eqref{hierarchy}, and then select $K_3$ such that $C_H=0$. After this, $K_1$ can be changed to the value that will give the desired value of $k_1$ by using the equation
(see, e.g., \cite{Wilson-Ewing:2017vju})
\begin{equation}  \label{choosingk1}
k_i = \frac{p_i K_i}{\Sigma_j p_j K_j}, \qquad \Rightarrow \quad
K_1 = \frac{k_1 (p_2 K_2 + p_3 K_3)}{(1 - k_1) p_1},
\end{equation}
which holds for classical Bianchi~I spacetimes (i.e., when spatial curvature and LQC are negligible, as they are when the initial conditions are set).

Selecting $K_1$ in this way will typically violate the Hamiltonian constraint, $C_H \neq 0$, so then $K_3$ must be updated, after which $K_1$ must in turn be modified again following \eqref{choosingk1} to get the correct Kasner exponent.  This process can be iterated, and in practice often converges to a set of initial conditions that both satisfies $C_H=0$ and gives the desired Kasner exponent $k_1$. Note however that sometimes this method fails, for example due to the iterated values of $K_1$ and/or $K_3$ no longer satisfying the inequalities \eqref{hierarchy}, in which case the process has to be restarted with different initial values for $p_i$ and $K_i$.

\subsection{Dynamics}

Given initial conditions derived following the procedure described above, we found numerical solutions to the six coupled ODEs \eqref{EqnsofMotion} for $(p_i, K_i)$, setting the lapse $N = \sqrt{p_1 p_2 p_3}$, and using the Butcher-1 sixth-order Runge-Kutta method \cite{Butcher:1964} with a fixed timestep, coded in C++.

%

There is a simple change of variables that significantly improves the numerical solver.  Before an LQC bounce when LQC effects are negligible, the $\gamma \bar{\mu}_i K_i$ values are very small and close to zero, and after the LQC bounce these terms tend to $-\pi$ as LQC effects decrease.  However, computers can handle floating point numbers with much higher precision when they are close to zero than when they are close to $-\pi$ (or any other non-zero number).  Therefore, after an LQC bounce the numerical error can accumulate rapidly leading the calculation to fail.  To avoid this, it is possible to shift the values of all three $K_i$ after the LQC bounce by
\begin{equation}
\gamma \bar\mu_i K_i \to \gamma \bar{\mu}_i K_i^{\rm new} = \pi + \gamma \bar{\mu}_i K_i,
\end{equation}
leaving the $p_i$ unchanged.  It is easy to verify that this is a discrete symmetry of the LQC effective theory.  In the Bianchi~IX spacetime, where there are multiple LQC bounces and recollapses, this redefinition is done after each LQC bounce.

To obtain a measure of the numerical error, we calculate the relative error
\begin{equation} \label{Rel_Error}
\mathcal{E}_r = \frac{|C_H|}{\sqrt{C_{H1,0}^2 + C_{H2,0}^2 + C_{H3,0}^2 + U_{i,0}^2}}.
\end{equation}
The subscript $0$ on the terms in the denominator indicate that these are the values from the initial condition---it is only $C_H$ in the numerator which is updated throughout the run.  Since the numerics use double floats, which have between 15-16 digits of precision, we expect the relative error to have a minimal possible value between $10^{-15}$ and $10^{-16}$.

Finally, we also developed a high-precision code that, although it required a considerably longer runtime, could be used to explore possible chaotic dynamics in the Bianchi~IX spacetime in Sec.~\ref{s.bIX}. This code was written in Julia with variable precision to 128 bits, and used the fourth-order Runge-Kutta method with a variable timestep to numerically determine the coupled ODEs for the Bianchi~IX LQC effective dynamics. In this code, for each run the timestep was manually set to appropriate values for different intervals of time to ensure the relative error $\mathcal{E}_r$ remained sufficiently small (we also tried several variable timestep codes where the timestep is automatically updated at each iteration, but we found that, at least for the codes we tried, they were not sufficiently accurate for our purposes).  Again, to measure the numerical error we calculated the relative departure of the Hamiltonian constraint from zero as defined in \eqref{Rel_Error}. This was used to determine the required timestep for each time interval: for each set of initial conditions, we ran the code several times and shortened the step size in any intervals where $\mathcal{E}_r$ would otherwise start to grow. In this way it was possible to find numerical solutions to the effective dynamics with a very low relative error $\mathcal{E}_r$ for long times.

\begin{figure}[t]
\begin{center}
\includegraphics[width=11cm]{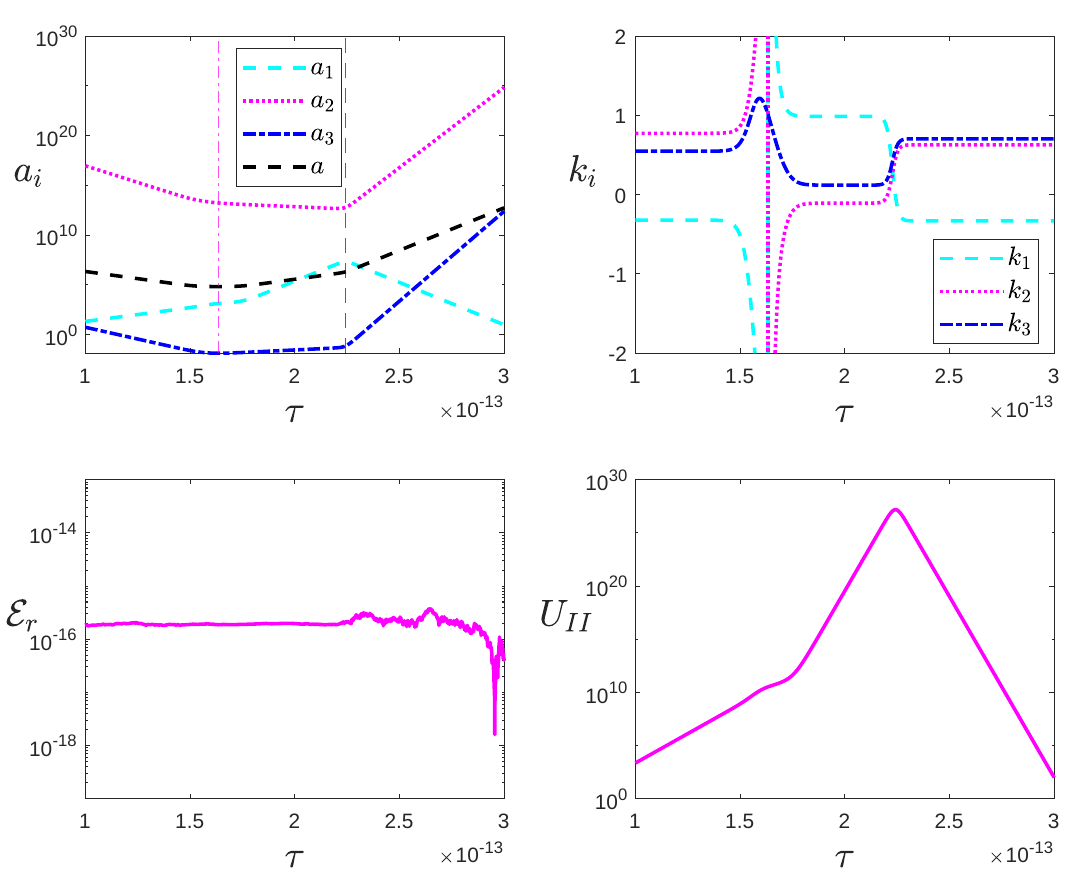}
\caption{This is an example of a Bianchi~II spacetime in the LQC effective theory, where the LQC bounce occurs before a Kasner transition. The upper left plot shows the directional scale factors as well as magenta and blue vertical lines showing the locations of the LQC bounce and of the Kasner transition, respectively, and the upper right plot shows the Kasner exponents. The lower plots show the relative error $\mathcal{E}_r$ and the value of the potential $U_{\!I\!I}$. The initial Kasner exponents are $(-0.320427, 0.773262, 0.547172)$, while the final Kasner exponents are $(-0.331894, 0.627930, 0.703944)$.
}
\label{f.b2a}
\end{center}
\end{figure}

\section{LQC Effective Dynamics: Bianchi II}
\label{s.bII}

In a vacuum Bianchi~II spacetime there are three possibilities for the LQC effective dynamics: either the LQC bounce occurs first and a Kasner transition occurs afterwards (this is possible for any initial value of $k_1$), or a Kasner transition occurs first, is followed by an LQC bounce, and either there is zero (if the initial $-\tfrac{1}{3} < k_1 < -\tfrac{2}{7}$) or one (if the initial $-\tfrac{2}{7} < k_1 < 0$) additional Kasner transitions after bounce. (Recall that the initial value of $k_1$ must lie in the range $-\tfrac{1}{3} < k_1 < 0$.)

\begin{figure}[t]
\begin{center}
\includegraphics[width=11cm]{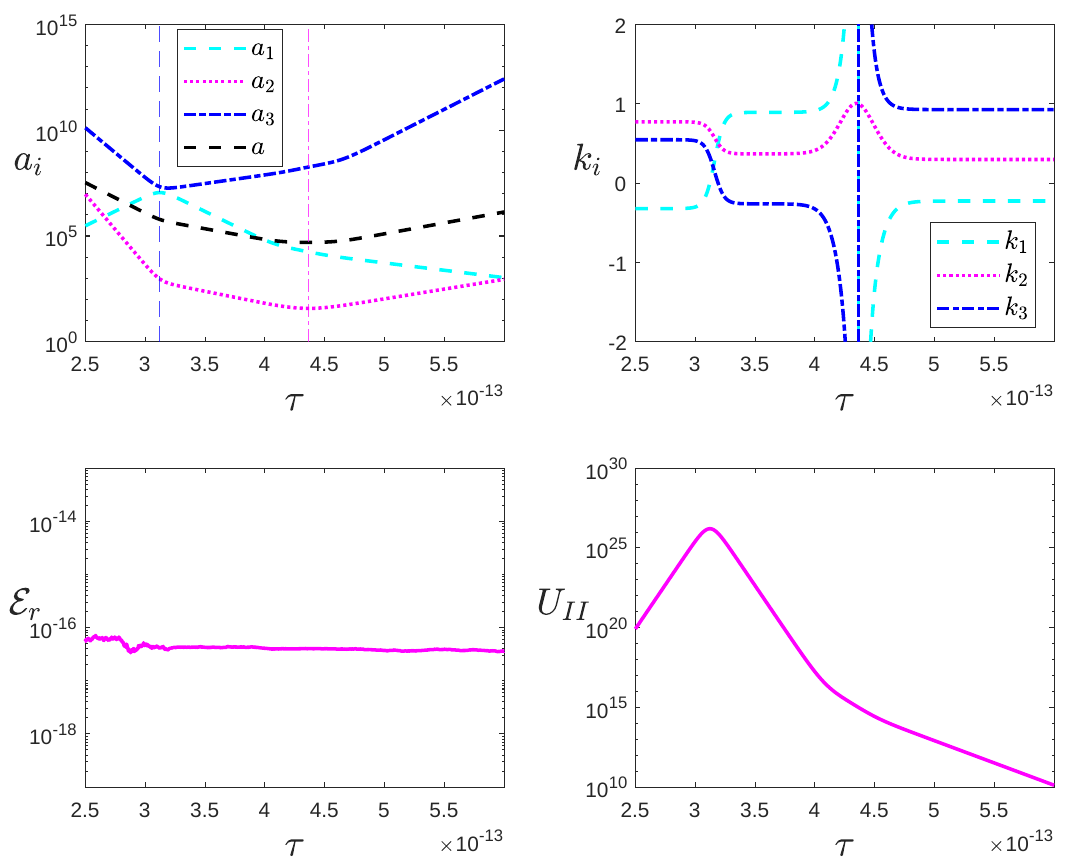}
\caption{This is an example of a Bianchi~II spacetime in the LQC effective theory, where a Kasner transition occurs before the LQC bounce. The upper left plot shows the directional scale factors as well as magenta and blue vertical lines showing the locations of the LQC bounce and of the Kasner transition, respectively, and the upper right plot shows the Kasner exponents. The lower plots show the relative error $\mathcal{E}_r$ and the value of the potential $U_{\!I\!I}$. The initial Kasner exponents are $(-0.320427, 0.773262, 0.547172)$, while the final Kasner exponents are $(-0.225544, 0.298010, 0.927533)$.
}
\label{f.b2b}
\end{center}
\end{figure}

\subsection{Clearly Separated Transitions}

\begin{figure}[t]
\begin{center}
\includegraphics[width=11cm]{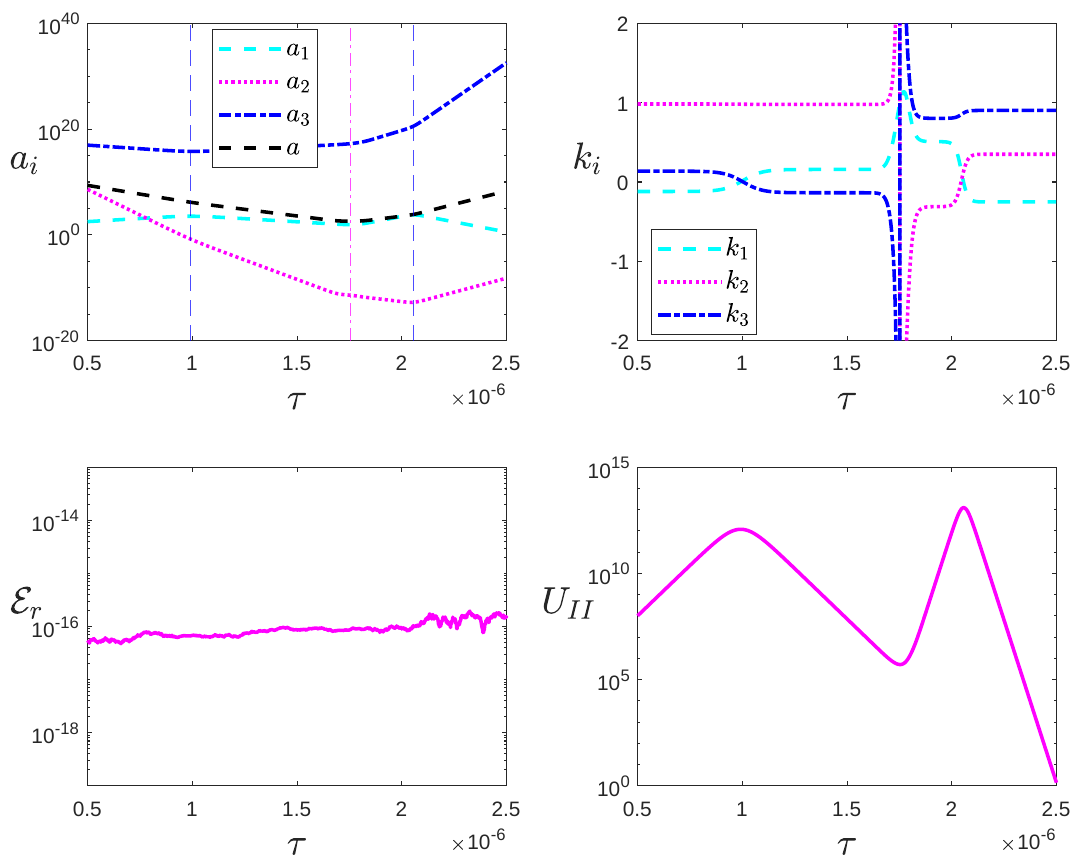}
\caption{This is an example of a Bianchi~II spacetime in the LQC effective theory, where there occurs a Kasner transition either side of the LQC bounce. The upper left plot shows the directional scale factors as well as magenta and blue vertical lines showing the locations of the LQC bounce and of the Kasner transitions, respectively, and the upper right plot shows the Kasner exponents. The lower plots show the relative error $\mathcal{E}_r$ and the value of the potential $U_{\!I\!I}$. The initial Kasner exponents are $(-0.119999, 0.983335, 0.136679)$, while the final Kasner exponents are $(-0.252179, 0.350008, 0.902148)$.
}
\label{f.b2c}
\end{center}
\end{figure}

In all cases, numerics show that if the Kasner transition(s) and LQC bounce are well separated, then the transformation rules for the Kasner exponents during Kasner transitions \eqref{kasner} and the LQC bounce \eqref{lqc-kas} hold to an excellent approximation.  Three representative examples, one for each of the cases described above, are shown in Figs.~\ref{f.b2a}--\ref{f.b2c}; in all cases the transition rules for the Kasner exponents hold to a better accuracy than one part in $10^4$.

In Fig.~\ref{f.b2a}, the first transition is due to the LQC bounce, and the second is a Kasner transition. For this particular run, the initial Kasner exponents are $(-0.320427, 0.773262, 0.547172)$, and the final Kasner exponents were found to be $(-0.331894, 0.627930, 0.703944)$; this can be compared with the predicted final Kasner exponents of $(-0.331887, 0.627933, 0.703951)$, calculated by applying the transformation rules \eqref{lqc-kas} first and then \eqref{kasner} to the initial Kasner exponents. The maximal relative error of these Kasner exponents corresponds to $\sim 2 \times 10^{-5}$.

In Fig.~\ref{f.b2b}, the initial Kasner exponents are $(-0.320427, 0.773262, 0.547172)$, these are exactly the same as for the example shown in Fig.~\ref{f.b2a}, except that in this case the Kasner transition occurs first, before the LQC bounce---the opposite order as compared to the example shown in Fig.~\ref{f.b2a}. The final Kasner exponents are $(-0.225544, 0.298010, 0.927533)$, as compared to the values $(-0.225527, 0.297992, 0.927517)$ predicted by applying first \eqref{kasner} and then \eqref{lqc-kas} to the initial Kasner exponents; the greatest relative error for these $k_i$ is $\sim 8 \times 10^{-5}$.

Finally, Fig.~\ref{f.b2c} shows an example where there is a Kasner transition either side of the LQC bounce. The initial Kasner exponents are $(-0.119999, 0.983335, 0.136679)$, and the final Kasner exponents are $(-0.252179, 0.350008, 0.902148)$. Comparing these to the predicted final Kasner exponents $(-0.252174, 0.350000, 0.902164)$ obtained by applying first \eqref{kasner}, then \eqref{lqc-kas}, and finally \eqref{kasner} again to the initial Kasner exponents, it is clear that there is once again an excellent match between the numerical results and the analytic transformation rules with the largest relative error being $\sim 2 \times 10^{-5}$.

We have presented here three representative examples of numerical solutions to the LQC effective dynamics for the Bianchi~II spacetime, considering the case where the Kasner transition(s) and the LQC bounce are well separated. These examples, as well as numerous other numerical solutions with various initial Kasner exponents, all show that the transition rules \eqref{kasner} and \eqref{lqc-kas} are extremely accurate when the Kasner transition(s) and LQC bounce are well separated.

\subsection{Nearly Simultaneous Transitions}

\begin{figure}[t]
\begin{center}
\includegraphics[width=11cm]{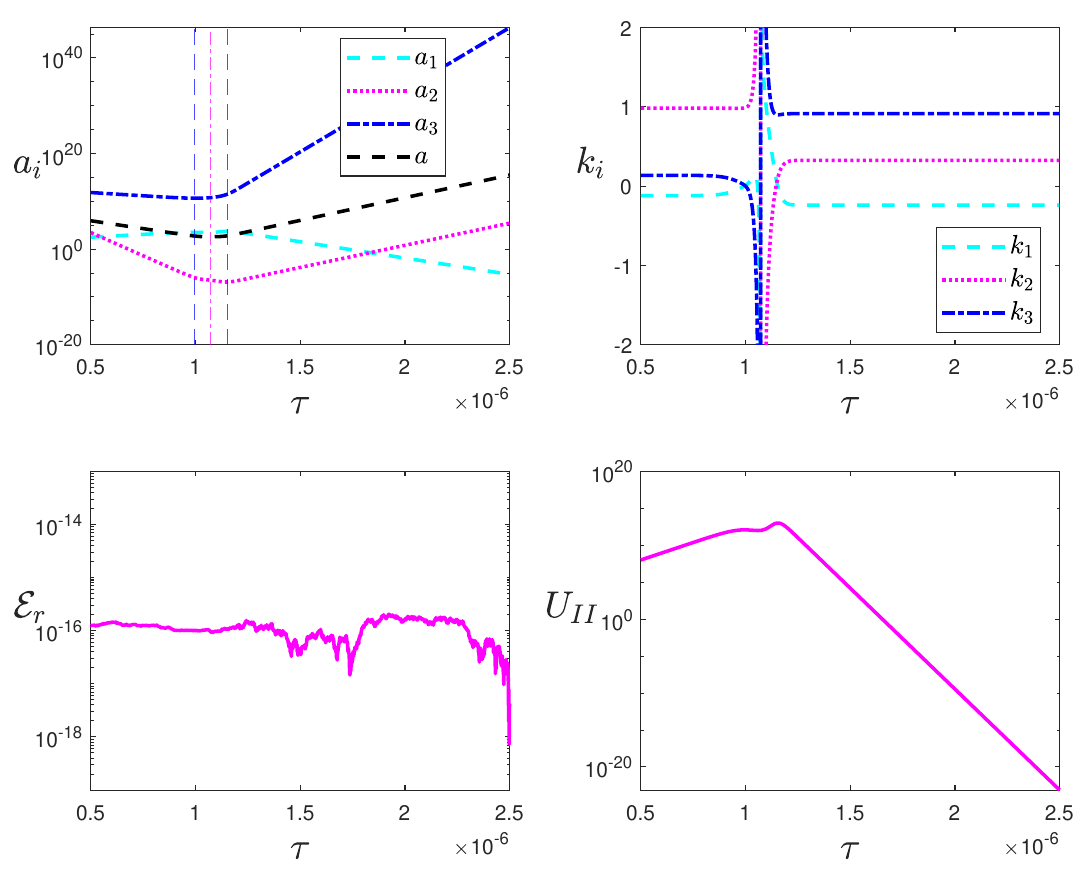}
\caption{
This is an example of a Bianchi~II spacetime in the LQC effective theory, where there occurs a Kasner transition either side of the LQC bounce but with very little time separating the transitions. The upper left plot shows the directional scale factors as well as magenta and blue vertical lines showing the locations of the LQC bounce and of the Kasner transitions, respectively, and the upper right plot shows the Kasner exponents. The lower plots show the relative error $\mathcal{E}_r$ and the value of the potential $U_{\!I\!I}$.
The initial Kasner exponents are $(-0.119999, 0.983335, 0.136679)$, while the final Kasner exponents are $(-0.240147, 0.325660, 0.914465)$.
}
\label{f.b2sim}
\end{center}
\end{figure}

The Kasner exponents transform in a more complicated fashion if both spatial curvature and LQC effects are simultaneously large; in this case numerics are needed to determine how the $k_i$ change, and the transition rules \eqref{kasner} and \eqref{lqc-kas} are no longer accurate. A representative example (with one Kasner transition before the LQC bounce, and another after) is shown in Fig.~\ref{f.b2sim}.

To understand the interplay between LQC and spatial curvature effects, it is useful to perform several calculations that all have the same initial Kasner exponents but with varying time intervals between the LQC bounce and the Kasner transition(s).  To do this, we modify the initial conditions so that when the initial conditions are set, the spatial curvature is unchanged, but the amplitude of LQC effects is increased or decreased.  Specifically, we change the initial conditions for the $p_i$ as $p_i \to p_i + d_i$; for simplicity we assume $d_2 = d_3$, while $d_1$ is chosen by requiring that $U_{\!I\!I}$ be unchanged.  Then, to ensure $C_H = 0$ we change the initial conditions for $K_i$ such that for each $i$ the combination $p_i K_i$ is left unchanged.  In this way, we can make LQC effects arise earlier or later in the numerics, while ensuring the spatial curvature becomes large at the same time, and keeping the same initial Kasner exponents.

\begin{figure}[t]
\begin{center}
\includegraphics[width=14cm]{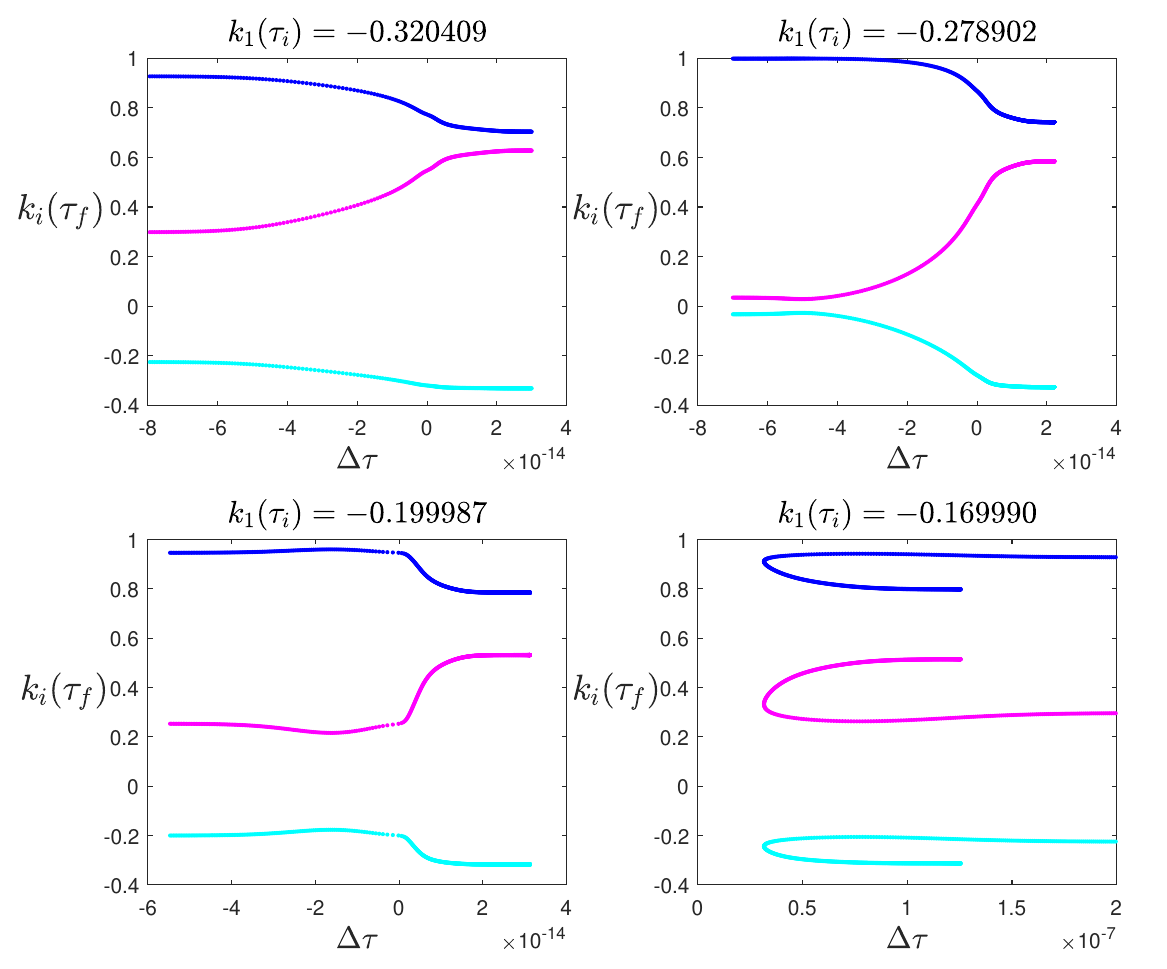}
\caption{Each plot shows a series of runs with fixed initial Kasner exponents and varying $\Delta \tau$. The bottom cyan line, the middle magenta line, and the top blue line correspond to $k_1(\tau_f)$, $k_2(\tau_f)$, and $k_3(\tau_f)$ respectively, and each group of three dots with the same $\Delta \tau$ show the values of $k_i(\tau_f)$ from an individual run (like the example shown in Fig.~\ref{f.b2sim}). For large $|\Delta \tau|$, the $k_i(\tau_f)$ asymptote to the value predicted by the appropriate combination of transformation rules, and there is a continuous transition between these values for smaller $\Delta \tau$.}
\label{f.move}
\end{center}
\end{figure}

For a given choice of initial Kasner exponents, we denote by $\Delta \tau$ the time interval between the LQC bounce and the Kasner transition, and calculate how the final Kasner exponents depend on $\Delta \tau$. (If there are two Kasner transitions, we select the Kasner transition where the potential $U_{\!I\!I}$ is maximized to calculate $\Delta \tau$. For the example shown in Fig.~\ref{f.b2c}, $\Delta \tau$ would be computed between the LQC bounce and the second Kasner transition where $U_{\!I\!I}$ reaches a greater maximal value than during the pre-bounce Kasner transition.)  We show four examples of a series of such runs in Fig.~\ref{f.move}.

For these four series of runs, each set of Kasner exponents corresponds to an entire run, giving the final Kasner exponents as a function of $\Delta \tau$.  From one run to another, the initial conditions are modified as described above so that the Kasner transition starts on one side of the bounce, and then slowly moves through the bounce and eventually occurs on the other side of the LQC bounce.  (Note that in some cases, there will be a second Kasner transition that will arise at some point, and $\Delta \tau$ may always have the same sign if the Kasner transition that maximizes $U_{\!I\!I}$ always remains on the same side of the potential; an example of this is shown in the bottom right plot of Fig.~\ref{f.move}.)

When $|\Delta \tau|$ becomes sufficiently large, then the transition rules \eqref{kasner} and \eqref{lqc-kas} can be used since (for sufficiently large $|\Delta \tau|$) the spatial curvature and LQC effects are not simultaneously large.  This is why the values of the final Kasner exponents asymptote to constant values for large $|\Delta \tau|$: in this case, the final values of the Kasner exponents is simply given by the appropriate combination of the transformation rules \eqref{kasner} and \eqref{lqc-kas}.  Note that there are two different asymptotic values for each run, since the transformation rules for a Kasner transition and an LQC bounce do not commute.

On the other hand, when $|\Delta \tau|$ is sufficiently small, the final values of the Kasner exponents typically lie between the two asymptotic values (although in this intermediate regime the final Kasner exponents do not evolve linearly with $\Delta \tau$).  Note that in some cases, like the bottom right plot in Fig.~\ref{f.move}, the final Kasner exponents will not always lie between the asymptotic values, but even in such cases the Kasner exponents do not differ significantly from the asymptotic values.

\begin{figure}[t]
\begin{center}
\includegraphics[width=11cm]{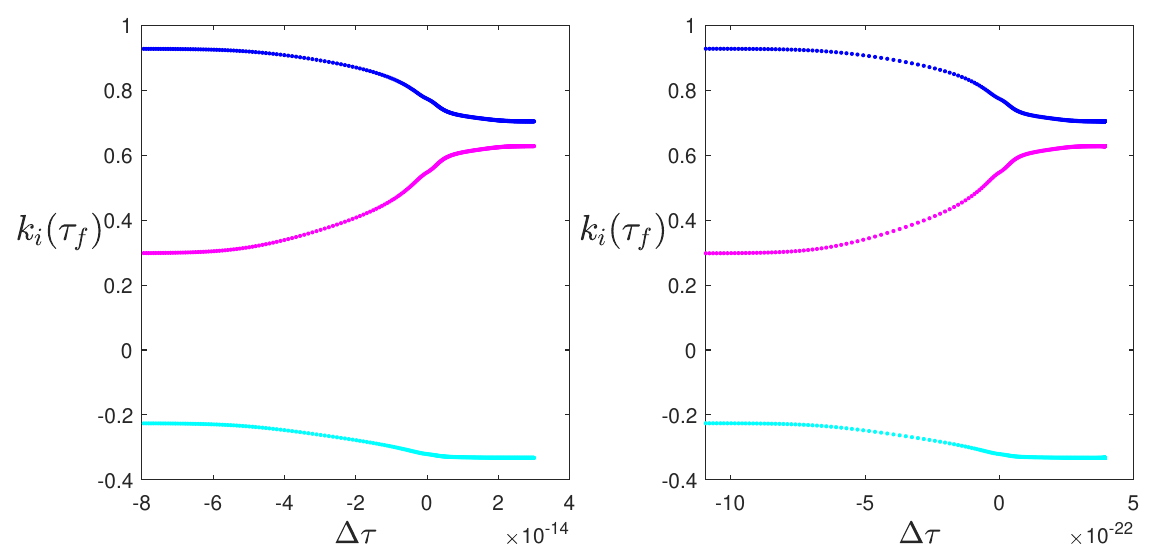}
\caption{The shape of the plot of $k_i(\tau_f)$ as a function of $\Delta \tau$ remains the same if the initial $p_i c_i$ are rescaled, up to an overall rescaling of $\Delta \tau$ by the inverse of that factor. In this example, $k_1(\tau_i) = -0.3204$ and the factor of rescaling is $1.3 \times 10^{-8}$ for each of the $p_i c_i$ terms, while $\Delta \tau$ is rescaled by the inverse factor $7.9 \times 10^7$.
}
\label{f.scale}
\end{center}
\end{figure}

An important point is that the shape of the $\bar{k}_i$ versus $\Delta \tau$ plots has a simple scaling symmetry: in the initial conditions, if each of the products $p_1 c_1$, $p_2 c_2$ and $p_3 c_3$ is rescaled by some factor $C$, then the initial Kasner exponents remain unchanged and the shape of the plot remains identical once $\Delta \tau$ is rescaled by the inverse factor $C^{-1}$, a representative example of this is shown in Fig.~\ref{f.scale}. The existence of this symmetry is due to the fact that the physics, once the Kasner exponents are fixed, depends on the characteristic timescale set by $(p_i c_i)^{-1}$; if this is rescaled, then so is $\Delta \tau$ by the same factor.

Finally, there are two interesting properties of $\Delta \tau$ that are worth pointing out.

First, note $\Delta \tau$ always remains positive only in the bottom right subplot of Fig.~\ref{f.move} where $k_1(\tau_i) = -0.169990$.  This is because when the LQC bounce occurs first then $\Delta \tau > 0$ always, while if there are Kasner transitions both before and after the bounce, then if initially $-2/7 < k_1 < -1/5$ it is the first (pre-bounce) spike in the potential that will be larger (and $\Delta \tau < 0$), while if initially $-1/5 < k_1 < 0$ it is the second (post-bounce) spike in the potential that will be larger (and $\Delta \tau > 0$); the bottom right subplot falls into this second category. This transition occurs at $k_1(\tau_i) = -1/5$, because this is the initial value of the Kasner exponents such that the rate of growth in $a_1$ is equal and opposite to itself before and after the LQC bounce, implying that the potential will grow equally rapidly either side of the LQC bounce.  To see why this is the case, we denote the sequence of Kasner exponents $(k_1^{(0)}, k_1^{(1)}, k_1^{(2)}, k_1^{(3)})$ corresponding to the initial value, the value after the first Kasner transition, the value after the LQC bounce, and the value after the second Kasner transition, respectively. For the rate of growth in $a_1$ to be equal and opposite either side of the LQC bounce requires $k_1^{(1)} = k_1^{(2)}$, and given the transformation rule \eqref{lqc-kas} for an LQC bounce this implies $k_1^{(1)} = k_1^{(2)} = 1/3$, which in turn fixes $k_1^{(0)} = -1/5$ by the standard Kasner transition rule \eqref{kasner}.

\begin{figure}[t]
\begin{center}
\includegraphics[width=11cm]{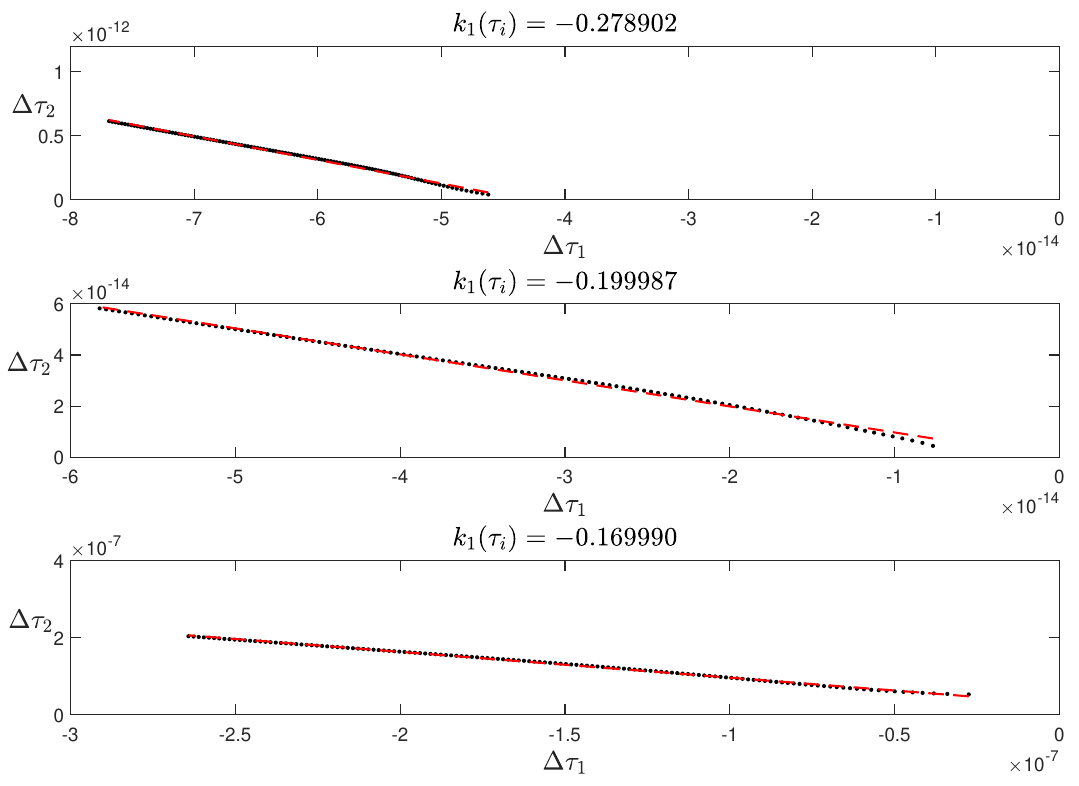}
\caption{These plots show the relation between $\Delta \tau_1$ and $\Delta \tau_2$ for three representative choices of initial $k_i$ in the case where there is a Kasner transition both before and after the LQC bounce. The black dots correspond to results from the numerical simulations, while the dashed red curve shows the best linear fit to the black dots. The slope is expected to be approximately given by (minus) the ratio between the value of the Kasner exponent $k_1$ before and after the LQC bounce; the numerical results match this expectation up to an error of $\sim 5\%$.}
\label{f.slopes}
\end{center}
\end{figure}

Second, for the case when there are two Kasner transitions it is possible to define $\Delta \tau_1$ and $\Delta \tau_2$ to be the time differences between the LQC bounce, and the first and second Kasner transitions respectively (note that with this definition, $\Delta \tau_1 < 0$ and $\Delta \tau_2 > 0$). With these definitions it is possible to compare how, for example, $\Delta \tau_2$ changes if $\Delta \tau_1$ is varied; three representative examples are shown in Fig.~\ref{f.slopes} and it is clear that for sufficiently large $|\Delta \tau_i|$ there is a linear relation between $\Delta \tau_1$ and $\Delta \tau_2$.

This relationship arises for a simple reason. For simplicity, in the following discussion we assume that the LQC bounce occurs at $\tau=0$, so the first Kasner transition occurs at $\tau = \Delta \tau_1$, and the second at $\tau = \Delta \tau_2$. For there to be a Kasner transition either side of the LQC bounce, it is necessary for the scale factor $a_1$ to increase before $\tau = \Delta \tau_1$, and also between $\tau=0$ and $\tau = \Delta \tau_2$. A Kasner transition occurs when $C_{Hi} \sim U_{\!I\!I} \sim a_1^4$, so if (for example) the initial conditions are modified so $|\Delta \tau_1|$ is increased and there is more time between the first Kasner transition and the LQC bounce, then $a_1$ will decrease by a larger amount between the first Kasner transition and the LQC bounce, and as a result more time will be required after the LQC bounce for $a_1$ to increase sufficiently to trigger the second Kasner transition.

More precisely, recall that $\ln a_1$ is linear in $\tau$, the time coordinate corresponding to $N = \sqrt{p_1 p_2 p_3}$ used in the numerics (see, e.g.,~\cite{Wilson-Ewing:2017vju} for details), so $\ln a_1$ decreases by an amount proportional to $|\Delta \tau_1|$ between the first Kasner transition and the LQC bounce, and increases by an amount proportional to $\Delta \tau_2$ between the LQC bounce and the second Kasner transition.  Since the Kasner transitions both occur at (approximately) the same value of $a_1$ (and therefore also $\ln a_1$), it follows that there must be a linear relation between $\Delta \tau_1$ and $\Delta \tau_2$, as is seen in Fig.~\ref{f.slopes}. Further, this calculation predicts that the slope is given, up to a minus sign, by the ratio between the values of the Kasner exponent $k_1$ before and after the LQC bounce. The results of the numerics agree with this prediction for the slope, although with an error of $\sim 5\%$. This error is likely due in part to the departures from the linear relation close to $\Delta \tau_i = 0$ that affect the measured slope, and in part to the approximation that the Kasner transition either side of the LQC bounce occurs at exactly the same value of $a_1$.

\section{LQC Effective Dynamics: Bianchi IX}
\label{s.bIX}

The dynamics of the Bianchi~IX spacetime are significantly more complex than for the Bianchi~I and II spacetimes since the potential $U_{\!I\!X}$, given in \eqref{Bianchi_Potentials}, forms a closed region and the system will `bounce' off the potential walls an infinite number of times. In addition, the spatial curvature in this case will cause a recollapse to occur in the mean scale factor; this combined with the LQC bounce gives a cyclic universe, with an infinite sequence of bounces and recollapses, though each cycle will typically be very different from the others.

Through these dynamics, the Kasner exponents behave in a more complicated manner near the recollapse, but away from the recollapse (and assuming the different transitions in the Kasner exponents are well separated) the Kasner exponents are expected to transform following \eqref{kasner} during Kasner transitions (with a cyclic permutation of the indices depending on which potential wall the system bounces off), and to transform as \eqref{lqc-kas} during an LQC bounce. These expectations are confirmed by the numerics.

\begin{figure}[t]
\begin{center}
\includegraphics[width=11cm]{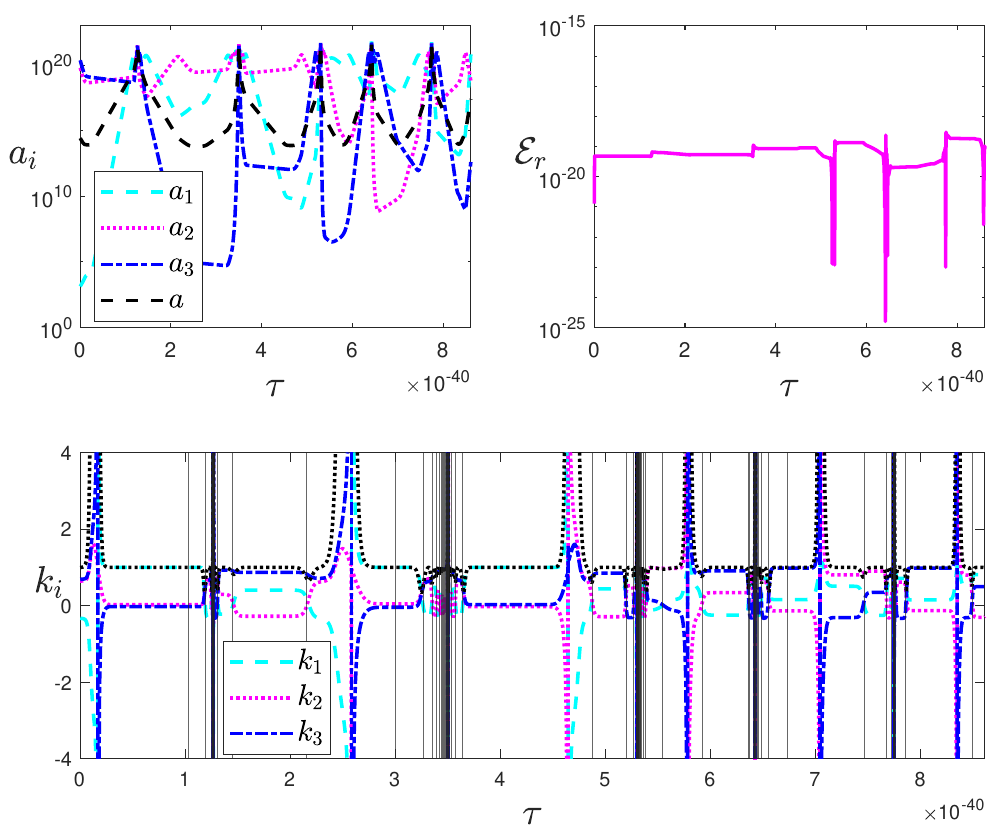}
\caption{
This is an example of a Bianchi~IX spacetime in the LQC effective theory.
The upper left plot shows the directional scale factors, and the upper right plot shows the relative error $\mathcal{E}_r$. The lower plot shows the Kasner exponents, with each vertical black line denoting the location of a Kasner transition; the LQC bounces are located where the Kasner exponents instantaneously diverge.
}
\label{b9-hp}
\end{center}
\end{figure}

\begin{figure}[t]
\begin{center}
\includegraphics[width=11cm]{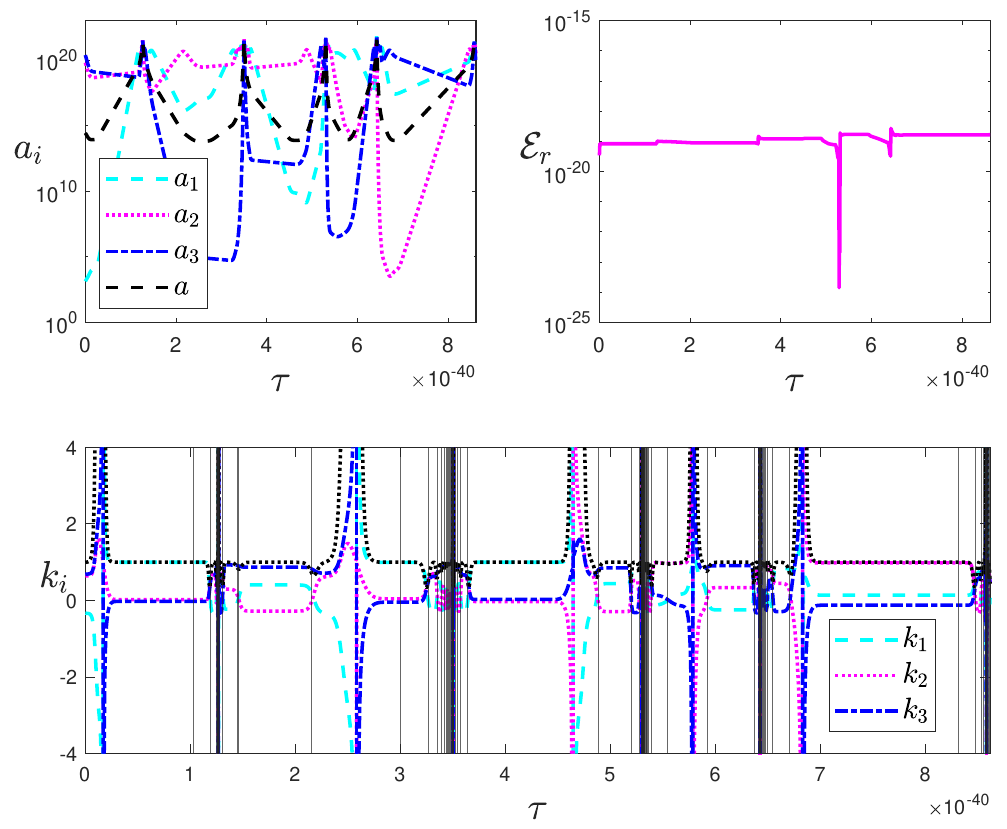}
\caption{
This is an example of a Bianchi~IX spacetime in the LQC effective theory, with initial conditions differing by 1 part in $10^{20}$ from the initial conditions for the solution shown in Fig.~\ref{b9-hp}. The upper left plot shows the directional scale factors, and the upper right plot shows the relative error $\mathcal{E}_r$. The lower plot shows the Kasner exponents, with each vertical black line denoting the location of a Kasner transition; the LQC bounces are located where the Kasner exponents instantaneously diverge.
}
\label{b9-hp_alt}
\end{center}
\end{figure}

A typical example is presented in Fig.~\ref{b9-hp}, with 6 cycles contained in the time interval shown for this example. In the bottom panel showing the Kasner exponents as a function of time, the location of the Kasner transitions (as defined by a local maximum of the potential $U_{\!I\!X}$ in time) is denoted by a black vertical line, while the LQC bounces are located where the Kasner exponents instantaneously diverge. It can be seen that there are many Kasner transitions close to the recollapse points, but relatively few near the LQC bounce. Further, it is interesting to note that there are relatively few Kasner transitions between any given LQC bounce and a neighbouring recollapse (outside the immediate vicinity of the recollapse where there are many Kasner transitions). This result agrees with earlier work that found there can only be a few Kasner transitions before the volume of the universe reaches $\sim \lp^3$ \cite{Doroshkevich}---and note that in our simulations the LQC bounce occurs when the spacetime curvature reaches the Planck scale, but the volume is still much larger than $\lp^3$ so there are even fewer Kasner transitions between the recollapse and the LQC bounce in this case. Note also that for this example a high-precision code was used to numerically determine the LQC effective dynamics for the Bianchi~IX spacetime, so the relative error $\mathcal{E}_r$ is considerably smaller here (at the price of a longer run time) than for the numerical solutions for the Bianchi~II spacetime presented in the previous section.

We have verified that when the spatial curvature is negligible during an LQC bounce, then the Kasner exponents transform following \eqref{lqc-kas}, as expected and exactly as was also found for Bianchi~II spacetimes as discussed in Sec.~\ref{s.bII}. However, in some cases the spatial curvature is important during the bounce and then this simple transformation rule no longer holds.  In general, there are three main possibilities: (a) the spatial curvature is entirely negligible, (b) only one term in the spatial curvature $U_{\!I\!X}$ is large during the bounce, or (c) two or more terms in the spatial curvature $U_{\!I\!X}$ are large during the bounce.  The second possibility can be subdivided into two subcases since if only one term in $U_{\!I\!X}$ becomes large then the situation is identical to a Bianchi~II spacetime (up to a cyclic permutation of the indices depending on which term in the potential is large), so in case (b1) it may be possible to predict the Kasner exponents after the LQC bounce using an appropriate combination of the Kasner transition \eqref{kasner} and LQC bounce \eqref{lqc-kas} transformation rules for the Kasner exponents if there is a slight separation between the LQC bounce and the Kasner transition(s), or on the other hand in case (b2) if the LQC bounce and Kasner transition(s) are sufficiently close then these transformation rules are not accurate. For each LQC bounce, we determine which of these four cases is realized by checking if the predicted (post-bounce) Kasner exponents agree with the numerical result within an accuracy of 1\%.

Based on the various numerical solutions we obtained, with a total of over 50 LQC bounces, we found that all four possibilities are approximately equally likely, although the case (a) was the most common at $\sim 32\%$, cases (b1) and (c) were approximately equally likely at $\sim 25\%$, while case (b2) was slightly less common at $\sim 18\%$. We leave a more detailed analysis of the relative frequency of these possibilities for future work.

\begin{figure}[t]
\begin{center}
\includegraphics[width=11cm]{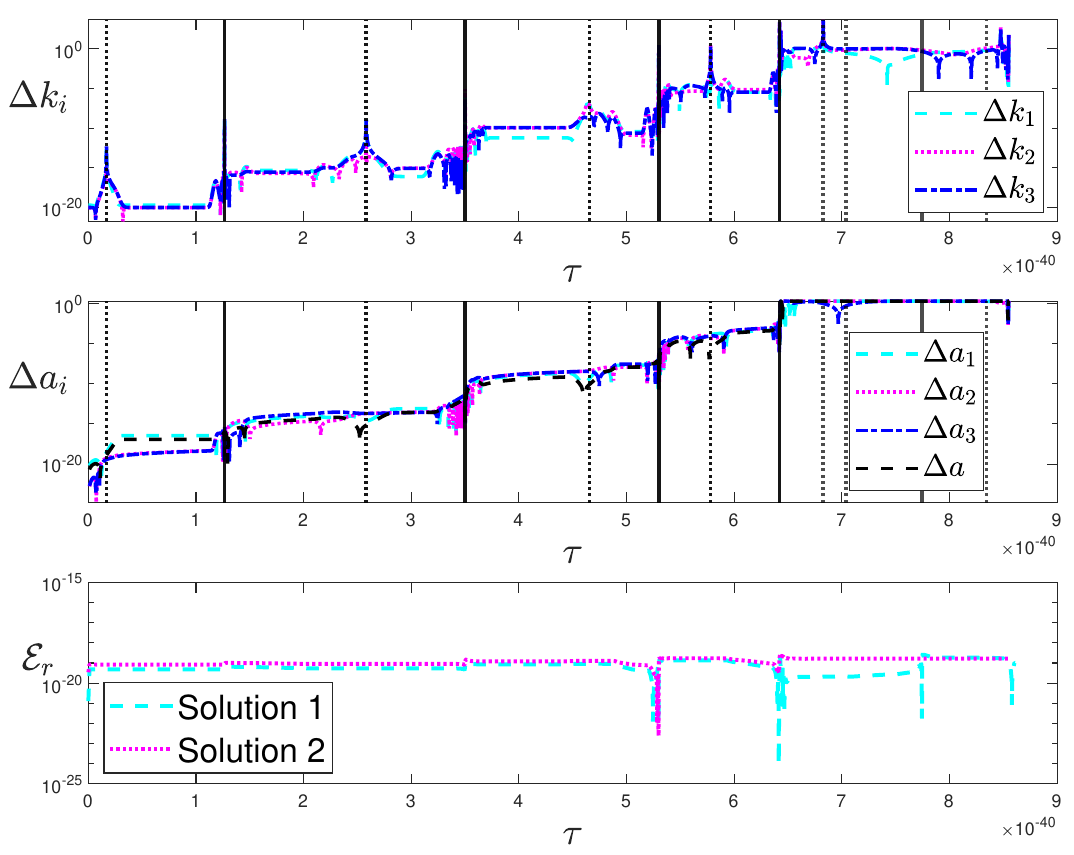}
\caption{
These plots show the differences $\Delta k_i$ (in the top plot) and $\Delta a_i$ (in the second plot) between the numerical solutions shown in Figs.~\ref{b9-hp} and \ref{b9-hp_alt}. The bottom plot shows the relative error $\mathcal{E}_r$ for the two runs. The dotted and solid vertical lines denote the locations of, respectively, LQC bounces and recollapses. As can be seen in the upper two plots, the two solutions diverge most strongly from each other around the points of recollapse.
}
\label{f.chaos}
\end{center}
\end{figure}

Since the classical dynamics for the Bianchi~IX spacetime are known to be chaotic \cite{Chernoff:1983zz, Cornish:1996yg}, it is interesting to consider the possibility that the Bianchi~IX LQC effective dynamics may be chaotic as well. Note that if this is indeed the case, then the chaos would have to arise in a somewhat different context. On the one hand, in the classical theory there is no bounce but there is an infinite number of Kasner transitions before the singularity is reached; the chaotic behaviour arises as the system approaches the singularity with the infinite number of Kasner transitions generating fractal boundary basins. On the other hand, in the LQC effective dynamics there is a non-singular bounce and the spacetime has an infinite number of bounce/recollapse cycles, with a finite number of Kasner transitions per cycle. This suggests that if there is to be chaos in the LQC effective theory it cannot arise in one cycle, but one must rather consider the infinite sequence of bounce-recollapse cycles.

In this work we do not conclusively determine whether the LQC effective dynamics for the Bianchi~IX spacetime are chaotic, but we do provide some evidence in its favour. In Fig.~\ref{b9-hp_alt} we show a numerical solution whose initial conditions differ by 1 part in $10^{20}$ from the initial conditions used to find the numerical solution shown in Fig.~\ref{b9-hp}. The differences between these two runs are shown in Fig.~\ref{f.chaos}, with
\be
\Delta a_i = 2 \left| \f{a_i - \tilde{a}_i}{a_i + \tilde{a}_i} \right|,
\ee
and similarly for $\Delta k_i$.

The plots in Fig.~\ref{f.chaos} clearly show that these solutions initially start extremely close, with $\Delta a_i \approx \Delta k_i \approx 10^{-20}$, and tend to diverge as $\tau$ increases. An interesting point is that the solutions tend to remain at approximately constant $\Delta a_i$ and $\Delta k_i$ during LQC bounces, and also in the period between an LQC bounce and a recollapse. On the other hand, the solutions diverge rapidly at the recollapse points where there is a rapid succession of Kasner transitions. We have seen the same qualitative behaviour for other pairs of numerical solutions that initially start close to each other.

These results suggest that the LQC effective dynamics for the Bianchi~IX spacetime are especially sensitive to initial conditions at the recollapse point: if there is chaos, then it is likely due to the classical dynamics around the recollapse, where quantum effects are entirely negligible. Rather, the contribution from quantum gravity is to cause the bounces, ensuring an infinite number of cycles, with an infinite number of recollapses that are each highly sensitive to the state of the system.

\section{Discussion}
\label{s.disc}

We have tested the transformation rules for the Kasner exponents close to the LQC bounce for the Bianchi type~II and type~IX spacetimes. When the spatial curvature is negligible during the bounce, then the transformation rule $k_i \to \tfrac{2}{3} - k_i$, derived for the Bianchi type~I spacetime \cite{Chiou:2007mg, Wilson-Ewing:2017vju}, holds to a high degree of accuracy, this is expected since when the spatial curvature is sufficiently small then the effective equations of motion for the Bianchi type~II and type~IX spacetimes are very well approximated by the Bianchi~I dynamics. On the other hand, if the spatial curvature is sufficiently large during the LQC bounce then the dynamics of the Bianchi~I spacetime (which has no spatial curvature) no longer provide a good approximation and the simple transition rule $k_i \to \tfrac{2}{3} - k_i$ does not hold in this case.  Instead, numerics are required to determine the full effective dynamics, through the LQC bounce, in order to calculate the impact of the spatial curvature on the change in the Kasner exponents; this was studied in detail for the Bianchi~II spacetime in Sec.~\ref{s.bII}. As can be expected due to its more complex dynamics, in the Bianchi~IX spacetime there are more ways for the spatial curvature to affect the transformation rules of the Kasner exponents---for example, two potential walls in $U_{\!I\!X}$ are relevant during $\sim 18\%$ of the LQC bounces, a case which cannot arise in the Bianchi~II spacetime where there is only one potential wall. On the other hand, even in the Bianchi~IX spacetime we found that during nearly a third of the LQC bounces the spatial curvature is entirely negligible and it is possible to use the $k_i \to \tfrac{2}{3} - k_i$ transformation rule with an accuracy better than $1\%$.

We also numerically explored the sensitivity of the LQC effective dynamics for the Bianchi~IX spacetime to the initial conditions, finding that nearby solutions separate rapidly during the recollapse phase. Combining this observation with previous work determining that the classical dynamics for this Bianchi~IX spacetime are chaotic \cite{Chernoff:1983zz, Cornish:1996yg}, it seems likely that this system is also chaotic. To test this, it will be necessary to look for chaos in a coordinate-independent fashion (in particular avoiding Lyapunov exponents which depend on the choice of the time coordinate), for example by determining whether the boundaries between different basins of attraction are fractal or not, as has been done for the classical dynamics \cite{Cornish:1996yg}. We leave a detailed investigation of this question for future work.

Finally, a natural question is to ask what ramifications these results may have on the picture suggested by the BKL conjecture, if the BKL conjecture is indeed correct. First, if neighbouring points decouple and the dynamics of each is that of an LQC Bianchi cosmology, then the singularity will be avoided by a non-singular bounce that occurs when the spacetime curvature reaches the Planck scale. Second, although the dynamics of the Bianchi~IX spacetime are sensitive to the initial conditions in a manner that is suggestive of chaos, it seems very unlikely that this sensitivity will have an impact on the BKL dynamics, since there will only be a finite number (of order $\sim 1$) of Kasner transitions near the LQC bounce, which will restrict how significantly the difference in the spacetime geometry at neighbouring points can grow. As was seen in the examples shown in Figs.~\ref{b9-hp}--\ref{f.chaos}, nearby solutions do not diverge especially strongly during the LQC bounce; instead, the system is more sensitive to the initial conditions during the recollapse.

As a final (and more speculative) comment concerning the BKL scenario in LQC, we point out that there could be two competing effects during the LQC bounce for an inhomogeneous spacetime: on the one hand, the sensitivity to the initial conditions may tend to increase the amplitude of the inhomogeneities, while on the other hand the repulsive effect due to quantum gravity that causes the bounce may also tend to suppress the inhomogeneities if the interactions between neighbouring points are not entirely suppressed. It would be interesting to quantify the strength of these two possible effects and to determine if one of the two is always the dominant effect in LQG, or if this depends on the initial conditions. A more complete understanding of LQG effects in inhomogeneous spacetimes (with potentially large inhomogeneities)---building on earlier work including a hybrid LQC/Fock quantization of the Gowdy spacetime \cite{Martin-Benito:2008eza, Garay:2010sk, MartinBenito:2010bh, Brizuela:2011ps} and the reformulation of the BKL dynamics in terms of connection variables \cite{Ashtekar:2008jb, Ashtekar:2011ck}---will be needed to answer this question.

\acknowledgments

\noindent
This work was supported in part by the Natural Sciences and Engineering Research Council of Canada. E.W.-E.~also acknowledges support from the UNB Fritz Grein Research Award.

\raggedright

\end{document}